\documentclass[12pt,tightenlines,eqsecnum,floats,aps,amsmath,amssymb,nofootinbib,prd]{revtex4}

\usepackage{hyperref}
\usepackage{comment}
\usepackage{color}

\usepackage{setspace}
\usepackage{amsmath,amssymb,amsfonts,amsthm,mathrsfs}
\usepackage{cancel}
\usepackage{graphicx}
\usepackage{enumerate}

\def\be{\begin{equation}}
\def\ee{\end{equation}}
\def\ba{\begin{eqnarray}}
\def\ea{\end{eqnarray}}
\def\bi{\begin{itemize}}
\def\ei{\end{itemize}}

\def\zb{\bar{z}}
\def\w{\omega}

\def\I{\mathcal{I}}
\def\L{\mathcal{L}}
\def\diff{\text{Diff}}
\def\tf{\text{TF}}
\def\w{\omega}

\def\hard{\text{hard}}
\def\soft{\text{soft}}
\def\Jo{\mathring{J}}
\def\mg{\text{mag}}
\def\gbms{\text{GBMS}}
\def\Db{\bar{D}}
\def\Rb{\bar{R}}
\def\alphab{\bar{\alpha}}

\def\O{\Omega}

\def\Nvac{N^\text{vac}}
\def\R{\mathcal{R}}

\def\stf{\text{STF}}
\def\here{\text{here}}
\def\there{\text{there}}

\newcommand\Nzero{\overset{\scriptscriptstyle 0}{N}\vphantom{N}}
\newcommand\None{\overset{\scriptscriptstyle 1}{N}\vphantom{N}}
\newcommand\Gpsi{\overset{\scriptscriptstyle \psi}{\Gamma}\vphantom{\Gamma}}
\newcommand\szero{\overset{\scriptscriptstyle 0}{S}\vphantom{S}}
\newcommand\sone{\overset{\scriptscriptstyle 1}{S}\vphantom{S}}
\newcommand\Xone{\overset{\scriptscriptstyle 1}{X}\vphantom{X}}
\newcommand\Rzero{\overset{\scriptscriptstyle 0}{R}\vphantom{R}}

\newcommand\Rone{\overset{\scriptscriptstyle 1}{R}\vphantom{R}}

\begin{document}
\title{Generalized BMS charge algebra}

\author{Miguel Campiglia}
\email{campi@fisica.edu.uy} 

\author{Javier Peraza}
\email{ jperaza@cmat.edu.uy} 
\affiliation{Facultad de Ciencias, Universidad de la Rep\'ublica \\ Igu\'a 4225, Montevideo, Uruguay}

\begin{abstract}
It has been argued that the symmetries of gravity at null infinity 
should include a Diff$(S^2)$ factor associated to diffeomorphisms on the  celestial sphere. However, the standard phase space of gravity does not support the action of such transformations. Building on earlier work by Laddha and one of the authors,  we present an extension of the phase space of gravity at null infinity on which Diff$(S^2)$ acts canonically.   The Poisson brackets of supertranslation  and  Diff$(S^2)$ charges reproduce the generalized BMS algebra introduced in  \cite{bms1}.  
\end{abstract}
\maketitle

\tableofcontents

\section{Introduction}
Since the work of Bondi, van der Burg, Metzner \cite{bms} and  Sachs \cite{sachs} it is known that the asymptotic  Killing symmetries of asymptotically flat spacetimes at null infinity are generated by supertranslations $\xi_f$ (labeled  by functions $f$ on the sphere) and Lorentz rotations $\xi_V$   (labeled by conformal Killing vector (CKV) fields  $V^a$ on the sphere), forming together the  BMS algebra
\be
[\xi_{f}, \xi_{f'} ]=0, \quad [\xi_V,\xi_f ] = \xi_{V(f)}, \quad   [\xi_V, \xi_{V'} ] = \xi_{[V, V']}.  \label{bmsalgebra0}
\ee
Here  $[V,V']$ is the Lie bracket of sphere vector fields and
\be
V(f) := V^a \partial_a f - \frac{1}{2}D_a V^a f.
\ee
This algebra can be  thought of as a generalization of the Poincare algebra, with translations 
replaced by the infinite dimensional abelian algebra of supertranslations.

As described by   Ashtekar and Streubel (AS)  \cite{aaprl,AS}, the gravitational field at null infinity has a natural phase space structure that allows  to associate  canonical charges  to  BMS symmetries.\footnote{In this paper, we use the word `charge' as a synonym of  canonical generator on the gravitational phase space at null infinity. This differs from other common usage of `charge' as a boundary term  associated to a  Cauchy slice  that ends at null infinity, see e.g. \cite{bunster}.} 
These are the  supermomenta $P_f$ and  angular momenta $\Jo_V$,  
with  Poisson brackets (PBs) reproducing the BMS algebra 
\be
\{P_{f}, P_{f'}\}=0, \quad \{\Jo_V,P_f\} = P_{V(f)}, \quad   \{ \Jo_V,\Jo_{V'} \} = \Jo_{[V, V']}  \label{bmsalgebra}. 
\ee

Many years after these foundational works, the subject of  gravitational  symmetries  at null infinity experienced two major revisions.  Firstly, Barnich and Troessaert (BT)  \cite{BTprl} studied an extension of the BMS algebra in which the vector fields $V^a$ are allowed to have poles, thus enlarging  the 6 dimensional algebra of global CKVs on the sphere into the infinite dimensional algebra of local CKVs. Whereas the BT extended BMS algebra has the same  form as (\ref{bmsalgebra0}), its associated charge algebra  exhibits an  extension term in the bracket between $P_f$ and $J_V$ \cite{BTcharge}. 
Secondly, Strominger and collaborators \cite{strom0,stromST} showed how BMS can be understood as a symmetry of the gravitational S-matrix, identifying  the corresponding supertranslation Ward identities with Weinberg's soft graviton theorem \cite{weinberg}. These two fronts came together in the work \cite{stromvirasoro}, where a subleading soft graviton factorization \cite{stromcach} was identified as a Ward identity  of BT superrotations.  Since then, there appeared many further developments  on  soft factorization and asymptotic symmetries, see  the reviews \cite{stromrev,acl,ruzz,pas} and references therein.

In \cite{bms1}, Laddha and one of the authors argued that, from the perspective of the subleading soft graviton theorem, it is more natural to consider a different generalization of the BMS algebra, one in which $V^a$ is allowed to be an arbitrary smooth  vector field on the sphere. The defining relations for this Generalized BMS (GBMS) algebra are again given by (\ref{bmsalgebra0}), and from the subleading soft graviton theorem one can identify a candidate for `super angular momentum' $J_V$.

It is then natural to ask if there exists an underlying phase space on which    $J_V$ acts.  A first step in this direction was taken in  \cite{bms2}, where an extension of the Ashtekar-Streubel phase space was identified and  $J_V$  obtained using covariant phase space techniques. The treatment in \cite{bms2}, however, had certain limitations  that   forbid an evaluation of PBs between charges. The objective  of the present work is to show that such limitations can be overcome. We will  obtain a phase space on which GBMS acts canonically with  Poisson brackets reproducing the GBMS algebra. \\

We proceed as follows (see next subsection for further details). Our starting point is the  observation that a Poisson bracket between $P_f$ and $J_V$ should satisfy 
\be
 \{J_V,P_f \} =   \delta_f J_V = - \delta_V P_f, \label{pbdeltas}
\ee
where $\delta_f$ and $\delta_V$ are the infinitesimal  transformations associated to supertranslations and superrotations respectively. 

It turns out that $\delta_V P_f$ can be evaluated from well-established expressions of supermomentum  
yielding
\be
\delta_V P_f = - P_{V(f)}, \label{delVP}
\ee
from which it follows   $J_V$ should satisfy 
\be
 \delta_f J_V =  P_{V(f)}. \label{delfJ}
\ee

Unfortunately, the  expression for $J_V$ given in \cite{bms1} does not satisfy (\ref{delfJ}) (as discussed later, this  is directly  related to the  non-closure of BT charges). 
Our main observation is that it is possible to correct  $J_V$ so that   (\ref{delfJ}) holds. The corrected $J_V$ 
is such  that (i) it reduces to the  angular momentum $\Jo_V$ when $V^a$ is  CKV and (ii) its Ward identity with the S matrix reproduces the subleading soft graviton theorem.\footnote{The old and new versions of $J_V$ lead to the same Ward identities if evaluated between finite energy states. To see their difference, one needs to evaluate Ward identities on states with zero energy gravitons \cite{akr} or   study the charge algebra on the S matrix \cite{dfh}. Both calculations require  double soft graviton formulas \cite{klose,arnab,chakra,bhatkar} and it is here where one can distinguish the new and old versions of $J_V$  \cite{extradouble}.}  
It thus satisfies the same conditions as the charge proposed in \cite{bms1}, with the advantage of being compatible with (\ref{pbdeltas}).

We will then verify the  corrected $J_V$  satisfies the remaining algebra relation, 
\be
\delta_V J_{V'} = -J_{[V,V']}. \label{delVJ}
\ee

Finally, we will see that conditions (\ref{delVP}), (\ref{delfJ}) and (\ref{delVJ}) can be used to determine an extension of the Ashtekar-Streubel phase space on which GBMS acts canonically, with   Poisson brackets   reproducing the GBMS algebra.    

A key technical tool we will rely upon  is a novel `superrotation-covariant' derivative  which   greatly facilitates  some of the computations. \\

We conclude the introduction by describing recent literature that relates to our work.

Compere, Fiorucci and Ruzziconi \cite{compere} improved the treatment in \cite{bms2} in several  directions, in particular by controlling radial divergences and defining renormalized surface charges.  Although their resulting GBMS charges and algebra are different from what we  find here (in particular they exhibit extension terms),   their analysis   was indispensable for the development of the present work.  

  Flanagan, Prabhu and Shehzad \cite{flanagan} present a no-go theorem for a symplectic structure supporting GBMS charges. 
Our symplectic structure violates at least one   the assumptions in their theorem and so there is in principle no contradiction with their result, see section  \ref{symplectic} for details.    Further subtleties in the construction of a phase space at null infinity are discussed in \cite{arzano,freidel}. 

   Adjei et.al. \cite{speranza} provide an interpretation of BT superrotations that leads to GBMS.   Potentially observable consequences of  superrotation charges are described in \cite{psz,nichols1,nichols2,nichols3,laddhasenobs,oliveri}. GBMS-like symmetries on null surfaces other than null infinity are discussed in \cite{penna1,penna2,cordovashao,cfp}.

\subsection{Strategy and outline}

Recall that gravitational radiation at null infinity is encoded in a 2d tensor $C_{ab}$ that captures the subleading (in $r$)  angular components of the spacetime metric. On the other hand, the leading angular components define a 2d metric $q_{ab}$  that is usually  kept fixed. From the perspective of GBMS, however, one needs to allow for variations of  $q_{ab}$ and it is here where    difficulties appear.

Let us for a moment forget about such difficulties and consider  Eq. (\ref{delfJ}), which, as argued, is a necessary condition for the existence of PBs. Since Eq. (\ref{delfJ})   does not involve variations of $q_{ab}$ (supertranslations do not change the 2-metric) we may try to solve it for $J_V$ with $q_{ab}$  given. The equation simplifies considerably  when  $q_{ab}$ is the unit round sphere  (referred to as `Bondi frame' \cite{aanullinf}), and so we  first focus on  this case in section  \ref{Jbondi}.

 To discuss Eqs. (\ref{delVP}) and (\ref{delVJ}) we need  expressions for $P_f$ and $J_V$ on a general $q_{ab}$. The former is well known in the literature, whereas the latter requires a generalization of the results in section \ref{Jbondi}. To do this generalization we revisit in section \ref{nonbondisec} the  description of non-Bondi frames. Using the  `superboost' field introduced in \cite{compere}, we define a `$\diff(S^2)$-covariant'  derivative that is covariant under the action of $\diff(S^2) \subset \gbms$ transformations. We will find that several non-Bondi frame formulas acquire a simple  geometrical meaning when written in terms of this derivative.  Some of the results in this section rely on appendices \ref{diffeos} and \ref{gerochapp}. 
 
  Using the tools of section \ref{nonbondisec}, in section \ref{JVsec} we obtain a general-frame formula for $J_V$ by `covariantizing' the expression obtained in section \ref{Jbondi}. We verify the resulting $J_V$ satisfies Eqs.    (\ref{delfJ}) and  (\ref{delVJ}).  Some of the results in this section rely on appendices \ref{Kidapp} and \ref{gbmsaction}.  
  
Finally, in section \ref{symplectic} we determine the symplectic structure on the space of pairs $(q_{ab},C_{ab})$ by demanding  compatibility  with the GBMS charges. \\

In the remainder of the section   we describe our conventions   and specify the assumed $|u| \to \infty$ fall-offs  at null infinity. We also present a brief review of GBMS transformations and   BMS charges.

\subsection{Conventions and spacetimes under consideration}
We work in units such that $32 \pi G =1$.  We consider asymptotically flat metrics at (future) null infinity in Bondi gauge (see e.g. \cite{BT}). The spacetime coordinates are given by a radial coordinate $r$, an advanced time  $u$, and angular coordinates $x^a$, $a=1,2$. The angular part of the spacetime metric has a $r \to \infty$ expansion of the form\footnote{We follow the parametrization used in \cite{compere} with  $(C_{ab})_\here= (\hat{C}_{AB})_\there$ and $(T_{ab})_\here=(N^{\text{vac}}_{AB})_\there$.}
\be
g_{ab}(r,u,x) \stackrel{r\to \infty}{=} r^2 q_{ab}(x) + r \left[ C_{ab}(u,x) + u T_{ab}(x) \right] + \cdots \label{gab}
\ee
The leading part of the angular metric, $q_{ab}$, is usually regarded as kinematical and fixed once and for all. The simplest choice, referred to as Bondi frame, is to take $q_{ab}$ the unit round sphere metric.
$C_{ab}$ satisfies $q^{ab}C_{ab}=0$ and encodes outgoing gravitational waves at future null infinity $\I$.   The tensor $T_{ab}(x)$ is constructed entirely from $q_{ab}$ and vanishes in  Bondi frame (see section \ref{nonbondisec} and appendix \ref{diffeos} for further details). 

We consider  $u \to \pm \infty$ fall-offs in $C_{ab}(u,x)$ compatible with a $O(1)$ subleading soft theorem (see e.g. \cite{laddhasen})
\be
\partial_u C_{ab}(u,x) \stackrel{u \to \pm \infty} = O(1/|u|^{2+\epsilon}), \quad \epsilon>0. \label{fallu}
\ee
These are compatible with tree-level scattering but are too restrictive for a  generic gravitational scattering where fall-offs are given by (\ref{fallu}) with $\epsilon=0$, corresponding to a logarithmic subleading soft theorem \cite{laddhasen,sahoosen}.\footnote{We thank Biswajit Sahoo for correcting a wrong statement in the first version of the manuscript.} 

We also require that $C_{ab}(u,x)$  is asymptotically flat as $u \to \pm \infty$. In Bondi frame, this corresponds to the vanishing of the magnetic part of  $C_{ab}(u,x)$ at  $u = \pm \infty$ (see e.g. \cite{strom0}):
\be
\lim_{u \to \pm \infty}  D_{[a} D^c C_{b]c}(u,x) =0. \label{magzero}
\ee
The non-Bondi frame version of (\ref{magzero}) is described  in  section \ref{magzerosec}.

GBMS  is generated by vector fields that preserve the Bondi form of the spacetime metric but are not necessarily 
Killing, thus allowing for changes in the leading order angular metric $q_{ab}$. The GBMS vector fields are parametrized by functions $f(x)$ (supertranslations) and arbitrary smooth vector fields $V^a(x)$ (which, borrowing the BT terminology,  we will call superrotations). They act on $q_{ab}$ and $C_{ab}$ according to  (see \cite{bms2} and section \ref{nonbondisec} for  details)
\be
\delta_f q_{ab}=0 ,    \quad \delta_V q_{ab} = \L_V q_{ab} - 2 \alpha q_{ab} \label{delqintro}
\ee
\be
 \delta_f C_{ab}  =  f \partial_u C_{ab} -2 D_a D_b f^\tf  + f T_{ab}  ,  \quad  \delta_V C_{ab} =  \L_V C_{ab}  + \alpha u \partial_u C_{ab} - \alpha C_{ab} , \label{delCintro}
\ee
where $D_a$ is the covariant derivative of $q_{ab}$,  $\tf$ stands for Trace-Free part, $\alpha=D_a V^a/2$, and $\L_V$ is the Lie derivative on the sphere. 

Note that the action of superrotations on the  2-metric $q_{ab}$ is such that it preserves the area element,  $\delta_V \sqrt{q}=0$. We will work in the space of metrics that can be reached  from Bondi-frame metrics by   finite GBMS transformations (see appendix \ref{diffeos}),  so that the area element of all $q_{ab}$'s  coincides with the unit round sphere area element.  In particular, $\alpha \equiv  \frac{1}{\sqrt{q}}\partial_a (\sqrt{q} V^a)/2$  is independent of $q_{ab}$. 

We conclude with a comment regarding the description of asymptotically flat spacetimes at null infinity. There are two main  approaches: The original due to Bondi and Sachs that we follow here, and the Penrose approach \cite{penrose} that uses a rescaled, compactified spacetime.  We expect the results presented here admit a  direct translation into the second description. See \cite{bms1,flanagan} for a discussion of GBMS in the Penrose approach.

\subsection{Review of BMS charges in Bondi frame}
In Bondi frame ($q_{ab}=$ unit sphere metric, $T_{ab}=0$) the asymptotic BMS Killing symmetries act on  $C_{ab}(u,x) $ according to
\ba
\delta_f C_{ab} & = &  f \partial_u C_{ab} -2 D_a D_b f^\tf  \label{delfo} , \\
\delta_V C_{ab} &=&  \L_V C_{ab}  + \alpha u \partial_u C_{ab} - \alpha C_{ab} , \label{delVo}
\ea
where $V^a$  are global CKVs of  $q_{ab}$ (i.e. they satisfy $\delta_V q_{ab}=0$).

As shown by Ashtekar and Streubel \cite{AS}, the transformations (\ref{delfo}) and (\ref{delVo}) are  canonical with respect to the  symplectic structure of gravitational radiation at null infinity,   
\be
\Omega= \int_{\I} du d^2 x \sqrt{q}(\delta \partial_u C^{ab} \wedge \delta C_{ab} ) . \label{OmegaAS}
\ee
This  allows one to find canonical charges associated to  BMS symmetries: the  supermomenta,
\be
P_f =   \int_{\I} du d^2 x \sqrt{q} \partial_u C^{ab} \delta_f C_{ab} \label{Pbms}
\ee
and  angular momenta
\be
\Jo_V  = \int_{\I} du d^2 x \sqrt{q} \partial_u C^{ab} \delta_V C_{ab}.\label{Jbms}
\ee
As mentioned in the introduction, these charges close under PBs, reproducing the BMS algebra.

\section{Super angular momentum in Bondi frame} \label{Jbondi}

The proposal  \cite{bms1,bms2}  for an asymptotic  GBMS symmetry provided the following candidate  for super angular momentum:\footnote{In \cite{bms2}, strong $u \to \pm \infty$ fall-offs where assumed such that the soft part of $J'_V$ was written as in (\ref{Jtilde}) with the replacement $u \partial_u C^{ab} \to - C^{ab}$. The soft charge as written here  was introduced in \cite{compere} and is valid for the more general fall-offs (\ref{fallu}).}
\be
J'_V = \int_{\I} du d^2 x \sqrt{q} \ \partial_u C^{ab} \delta_V C_{ab}  +  \int_{\I} du d^2 x \sqrt{q} \ u \partial_u C^{ab}(-4 D_a D_b \alpha + D_a D^c \delta_V q_{bc}- \delta_V q_{ab}), \label{Jtilde}
\ee
with $\delta_V C_{ab}$ and $ \delta_V q_{ab}$ given in Eqs. (\ref{delqintro}), (\ref{delCintro}).

Compared to the BMS angular momentum (\ref{Jbms}),  $J'_V$ has an extra `soft' term that vanishes when $V^a$ is a global CKV.  We recall from \cite{bms2} that  if $J'_V$ is written in terms of holomorphic coordinates on the sphere, the resulting expression coincides with the BT  charge used in \cite{stromvirasoro} to compute  BT superrotation Ward identities. 

As motivated in the introduction, it is of interest to evaluate the action of a supertranslation on $J'_V$. To organize  the calculation, we  introduce the following notation:\footnote{The notation is inspired from the role of  charges on soft theorems.  $\Nzero_{ab}(x)$ and  $\None_{ab}(x)$ are the leading and subleading soft modes \cite{2dstress} of the news tensor $N_{ab}(u,x) = \partial_u C_{ab}(u,x)$.}
\ba
\szero^f_{ab} & := & -2 D_a D_b f^\tf \\
\sone^V_{ab} & := &  [-4 D_a D_b \alpha + D_{(a} D^c \delta_V q_{b)c}- \delta_V q_{ab} ]^\tf\\
\Nzero_{ab}(x) & := & \int_{-\infty}^\infty  \partial_u C_{ab}(u,x) du \label{Nzero} \\
\None_{ab}(x) & := & \int_{-\infty}^\infty  u \partial_u C_{ab}(u,x) du , \label{None}
\ea
and write the supermomentum (\ref{Pbms}) and super angular momentum (\ref{Jtilde}) as
\ba
P_f & = &   P_f^\hard  + \int_{S^2} d^2 x \sqrt{q} \Nzero^{ab} \szero^f_{ab} , \\
J'_V & =& J^\hard_V+  \int_{S^2} d^2 x \sqrt{q} \None^{ab} \sone^V_{ab},
\ea
where the `hard' piece is the part of the charge that involves an integral over $\I$ of terms quadratic in $C_{ab}$. To evaluate $\delta_f J'_V$ we note the following identities
\ba
\delta_f J^\hard_V & =  &  P^\hard_{V(f)}  + \int_{S^2} d^2 x \sqrt{q} \Nzero^{ab} (\L_V - \alpha)\szero^f_{ab}  \\ 
\delta_f  \None_{ab} & = &- f  \Nzero_{ab} ,
\ea
from which we arrive at
\be
\delta_f J'_V = P_{V(f)} +K(f,V) \label{delfnc}
\ee
where
\be
K(f,V) =\int_{S^2} d^2 x \sqrt{q} \Nzero^{ab}(  (\L_V - \alpha)\szero^f_{ab} - \szero^{V(f)}_{ab} - f \sone^V_{ab}). \label{defK}
\ee
$K(f,V)$ may be thought of as a non-CKV generalization of the BT extension term \cite{BTcharge}; see subsection \ref{localCKV} for further comparison with  BT charges.

We now make use of a non-trivial identity, shown in appendix \ref{Kidapp}, which expresses  $K(f,V)$  as a total $\delta_f$ term plus a  `magnetic' piece,\footnote{We become aware of   identity (\ref{Kidentity}) from the  expressions of $\delta_f N_a$ in \cite{BT} ($N_a$ =   angular momentum aspect). The improved super angular momentum (\ref{correctedJ}) corresponds to $J_V = -4 \int du d^2 x \sqrt{q} V^a \partial_u N_a^{HPS}$ where $N_a^{HPS}$ is, modulo the soft piece,  the angular momentum aspect as defined in \cite{hps}. See \cite{compere} for a comparison of the different conventions for  $N_a$.}
\be
K(f,V) =-\delta_f  J_V^{\partial \I}  + \mg(f,V), \label{Kidentity}
\ee
where
\ba
J_V^{\partial \I}& =& \int_{\partial \I} d^2 x \sqrt{q} (V^a C^{bc} D_c C_{ab} + \frac{3}{2}\alpha C^{ab}C_{ab}) \label{Jextra0} \\
\mg(f,V) &=& -4 \int_{S^2} d^2 x \sqrt{q} \Nzero^{ab}D_a D^c(D_{[b}f V_{c]}- \frac{1}{2}f D_{[b}V_{c]})  =0 .\label{mag}
\ea
In (\ref{Jextra0})  $  \int_{\partial \I} \equiv   \int_{\partial \I_+} -   \int_{\partial \I_-}$ is a difference of integrals over the $u= \pm \infty$ boundaries of $\I$. The vanishing of (\ref{mag}) is  due to condition (\ref{magzero}) which implies
\be
D^{[c} D_{a} \Nzero^{b]a}=0. \label{magzero2}
\ee

Thus, if we  redefine  the super angular momentum as
\be
J_V = J'_V +J_V^{\partial \I} , \label{correctedJ}
\ee
it will satisfy
\be
\delta_f J_V = P_{V(f)}. \label{delfJP}
\ee
Let us make a few comments about the proposed expression  for  super angular momentum:
\begin{enumerate}
\item $J_V^{\partial \I}$  vanishes when $V^a$ is a global CKV thanks to condition (\ref{magzero2}). (This vanishing is not  obvious at first sight, see next subsection for an explicit demonstration). Thus, $J_V$ reduces to the standard BMS angular momentum $\Jo_V$ when $V^a$ is a global CKV.
\item One can formally write an operator expression  for $J_V^{\partial \I}$  in terms of graviton Fock operators \cite{extradouble}. The resulting expression has a trivial action on finite energy states and thus has no effect on the usual  computation of single-charge Ward identities  \cite{stromvirasoro,bms1}. It can, however, affect the evaluation of double-charge Ward identities \cite{dfh}, or single-charge Ward identity if one of the external states is in a shifted vacuum \cite{akr}. The consequences of such term on Ward identities will be discussed elsewhere  \cite{extradouble}.

\item Eq. (\ref{Kidentity}) defines $J_V^{\partial \I}$ modulo terms that are annihilated by $\delta_f$. Such terms can be constructed from the leading mode $\Nzero_{ab}$ and its powers. We discard such possible contributions since  (i) linear terms in $\Nzero_{ab}$ would spoil the S matrix Ward identities and (ii)  higher powers in $\Nzero_{ab}$ are non-local in $u$ (they cannot be written as an integral over $\I$ of a density local in $C_{ab}$).
\end{enumerate}

In the remainder of the section we will present an alternative form of $J_V^{\partial \I}$ that will be of later use.  We will also make contact with the BT treatment by describing how expressions simplify in the case of  local CKVs. 

\subsection{Alternative form of $J_V^{\partial \I}$}
Let
\be
C_{ab}^\pm(x) := \lim_{u \to \pm \infty} C_{ab}(u,x)
\ee
be the asymptotic  values of $C_{ab}$ at  $u = \pm \infty$. The  vanishing of their magnetic part (\ref{magzero}) implies they can be written as 
\be
C^\pm_{ab}= - 2 (D_a D_b C^{\pm})^\tf \equiv \szero_{ab}^{C^\pm} \label{stgboson}
\ee 
for some functions    $C^{\pm}(x)$.  
Below we use  (\ref{stgboson}) to provide an alternative expression for $J_V^{\partial \I}$. 

We start by writing the soft mode $\Nzero_{ab}$  as
\be
\Nzero_{ab} = C^+_{ab}- C^-_{ab}. \label{NzeroCpm}
\ee
Substituting (\ref{NzeroCpm}) in the definitions of $K(f,V)$ and  $J_V^{\partial \I}$  we find they can be written as 
\ba
K(f,V) &=  K^+(f,V) - K^-(f,V) \\
J_V^{\partial \I} & = J_V^{\partial \I_+}- J_V^{\partial \I_-} \label{Jpartialdiff}
\ea
where $\pm$ indicate the terms that depend on $C_{ab}^\pm$. 

We next observe that identity (\ref{Kidentity}) holds  for each piece separately:
\be
K^\pm(f,V) =-\delta_f  J_V^{\partial \I_\pm}  \label{Kidentity2}
\ee
($\mg^\pm(f,V) = 0$). Finally, since $J_V^{\partial \I_\pm}$ is quadratic in $C^{\pm}$ and
\be
\delta_f C^{\pm}  =  f  \label{delfC} 
\ee
we have
\be
J_V^{\partial \I_\pm} = \frac{1}{2} \delta_{f} J_V^{\partial \I_\pm}|_{f=C^\pm} \label{delfJVp} = - \frac{1}{2} K^\pm(C^{\pm},V), 
\ee
where in the last equality we used Eqs. (\ref{Kidentity2}).

Collecting the above results, we conclude  $J_V^{\partial \I}$ can be written as
\be
J_V^{\partial \I} = - \frac{1}{2} ( K^+(C^{+},V) - K^-(C^{-},V) ). \label{Jextra2}
\ee
This form makes it manifest that  $J_V^{\partial \I}$ vanishes for global CKV  (see also next subsection).

\subsection{Holomorphic coordinates and local CKVs.} \label{localCKV}
In holomorphic coordinates $(z,\zb)$ such that $q_{zz}=0=q_{\zb \zb}$ one has 
\ba
\sone^V_{zz} & = & -2 \partial^3_z V^z \\
 (\L_V - \alpha)\szero^f_{zz} - \szero^{V(f)}_{zz} & = & - 2 \delta_V (D^2_z)  f
\ea
and (\ref{Jextra2}) becomes
\be
J_V^{\partial \I} = \int_{\partial \I} d^2 z \sqrt{q} C^{z z} ( \delta_V(D^2_z)  C - C \partial^3_z V^z ) + c.c.
\ee
If $V^a$ is a global CKV then $\delta_V(D^2_z)=0 =\partial^3_z V^z$ and  $J_V^{\partial \I}$ vanishes as stated before. 


If  on the other hand $V^a$ is a  local CKV (i.e. $\partial_{\zb}V^z=0$)  one can show that
\be
 \delta_V(D^2_z) C = \frac{1}{2}\partial^3_z V^z C .
\ee
In this case  $J_V^{\partial \I}$ and $K(f,V)$ reduce to
\ba
J_V^{\partial \I} & =&  -\frac{1}{2}\int_{\partial \I} d^2 z  \sqrt{q} C^{z z} C \partial^3_z V^z ) + c.c. \label{Jpartial3} \\
K(f,V) & =& \int d^2 z \sqrt{q}  \Nzero^{z z} f \partial^3_z V^z + c.c.  \label{K3}
\ea
Expression (\ref{K3}) corresponds to the  BT extension as written in \cite{dfh}. We  see that $J_V^{\partial \I}$  is non-trivial for BT superrotations, and may also be used  to cancel the extension term as in  the smooth vector field case.

\section{Non-Bondi frames} \label{nonbondisec}
In the case where $q_{ab}$  is not round sphere metric, one needs an additional  $u$-independent tensor $T_{ab}$ to appropriately describe the gravitational field at null infinity \cite{geroch,compere}. This tensor was   introduced by Geroch \cite{geroch} in order to have a conformally invariant notion of gravitational radiation  in the Penrose description of asymptotically flat spacetimes.  Here, following \cite{compere}, we introduce $T_{ab}$ in the definition of  $C_{ab}$ as given in Eq. (\ref{gab}), 
\be
g_{ab} \stackrel{r\to \infty}{=} r^2 q_{ab} + r \left[ C_{ab} + u T_{ab} \right] + \cdots \label{gab2}
\ee
so that $C_{ab}=0$ represents a flat metric in a non-Bondi frame, see appendix \ref{diffeos} for details.

With this definition, the action of supertranslations and superrotations on $C_{ab}$ is given by
\ba
\delta_f C_{ab} & = &  f \partial_u C_{ab} -2 D_a D_b f^\tf  + f T_{ab} , \label{delf} \\
\delta_V C_{ab} &=&  \L_V C_{ab}  + \alpha u \partial_u C_{ab} - \alpha C_{ab} . \label{delV}
\ea
We see that  supertranslations acquire an extra term with respect to the Bondi-frame expression (\ref{delfo}).  The new expression  generalizes to non-Bondi frames the fact  that  the inhomogeneous piece of $\delta_f C_{ab}$ vanishes for spacetime translations \cite{geroch}:
\be
-2 D_a D_b f^\tf  + f T_{ab}  = 0 \iff f = \text{spacetime translation}.
\ee
Regarding superrotations (\ref{delV}), we note they lack an inhomogeneous term that appears with the usual definition of $C_{ab}$  \cite{BT,bms2}.   From this perspective, the role of $T_{ab}$ in (\ref{gab2}) is to eliminate such  inhomogeneous term, see  section 4 of \cite{bms2}.

In the original literature of  BMS in the  Penrose  approach, special care is taken to ensure   frame-independence, see e.g. \cite{geroch,jmp}.  In particular, the Ashtekar-Streubel  expression for supermomenta is  valid in any frame. When  written in terms of the physical spacetime metric (\ref{gab}), the AS supermomentum  takes the form
\be
P_f =   \int_{\I} du d^2 x \sqrt{q} \partial_u C^{ab} \delta_f C_{ab} \label{PfAS}
\ee
with $\delta_f$ given in (\ref{delf}).    Given the transformation rules of $C_{ab}$   and  $q_{ab}$  under superrotations, we can compute $\delta_V P_f$ resulting in  (see appendix \ref{gbmsaction})
\be
\delta_V P_f = -P_{V(f)} .\label{delVPf}
\ee

In the next section we will obtain a general-frame expression of super angular momentum  compatible with (\ref{delVPf}) in the sense of Eq. (\ref{pbdeltas}). A main tool we will use to this end is a `superrotation-covariant' derivative that can be defined with the help of a potential we now introduce.

\subsection{$\psi$-potential } \label{psipotential}
In \cite{compere}  a `superboost' field $\psi$ was introduced that serves as a   potential   for  $T_{ab}$  in the sense that\footnote{We are deviating from the notation in \cite{compere}: $\psi_{\text{here}}=- \Phi_{\text{there}}/2$.} 
\be
T_{ab}= 2 (D_a \psi D_b \psi + D_a D_b \psi)^\tf. \label{defT}
\ee
Under superrotations  $\psi$ transforms according to \cite{compere}
\be
\delta_V \psi = \L_V \psi - \alpha   , \label{delpsi}
\ee
which can be verified  to be compatible with the transformation  of  $T_{ab}$ induced by $\delta_V q_{ab}$\cite{bms2}
\be
\delta_V  T_{ab} = \L_V T_{ab} - 2 D_a D_b \alpha^\tf \label{delVT}.
\ee

One aspect of $\psi$ we would like to bring attention to is that, unlike $T_{ab}$, it is not invariant under CKVs.  In other words,  $q_{ab}$ fixes $\psi$ modulo  an ambiguity   parametrized by the conformal isometries of $q_{ab}$.\footnote{This ambiguity can be fixed by defining  $\psi$ with respect to a reference 2d metric from which all other $q_{ab}$'s are obtained by finite superrotations. Here we do so by considering a reference unit sphere metric, see  appendix \ref{diffeos}. Another natural choice   is to consider the Euclidean plane as a reference metric \cite{compere}.}  We will later see  that all quantities of interest such as charges and symplectic structure   depend on $\psi$ only  through the combination (\ref{defT}).  That is,   our expressions will in fact be  independent of the ambiguity in  $\psi$.  This property will not always be manifest and in some cases it will require some work to establish it.  Further details on $\psi$ and its relation with the Geroch tensor are given in  appendices  \ref{diffeos} and \ref{gerochapp}.  

We  now use   $\psi$ to construct a `superrotation-covariant'  derivative.

\subsection{$\diff(S^2)$-covariant derivative} \label{covderivative}

Let us first define the notion of covariance of a tensor with respect to superrotations. We say a ($u$-independent) tensor  $t_{a_1 \ldots}^{b_1 \ldots}$ on the celestial sphere is covariant under  superrotations if it satisfies the transformation rule,
\be
\delta_V t_{a_1\ldots}^{b_1\ldots} = \L_V t_{a_1\ldots}^{b_1\ldots} + k \alpha t_{a_1\ldots}^{b_1\ldots} \label{delVt}
\ee
for some constant $k$. For instance, the metric $q_{ab}$ is a covariant tensor with  $k=-2$. Other  examples are  the leading  and subleading  soft modes of the news tensor introduced  in Eqs. (\ref{Nzero}), (\ref{None}), whose  transformation properties obtained from  (\ref{delV}) are:
\ba
\delta_V \Nzero_{ab} &= & \L_V \Nzero_{ab} - \alpha \Nzero_{ab} \\
\delta_V \None_{ab}  &= & \L_V \None_{ab} - 2 \alpha \None_{ab}.
\ea
On the other hand, the potential $\psi$ and the tensor $T_{ab}$ are examples of non-covariant tensors.\\

Given a covariant tensor as defined above, its regular covariant derivative will not be covariant under superrotations. For example,  consider a scalar $\varphi$ such that $\delta_V \varphi = (\L_V + k \alpha)\varphi$. Then,
\be
\delta_V (D_a \varphi)  = D_a(\L_V \varphi +k \alpha \varphi) 
= (\L_V + k \alpha) D_a \varphi + k D_a \alpha  \varphi \neq  (\L_V + k \alpha) D_a \varphi.
\ee
The `extra' $k D_a \alpha \varphi$ term can  be canceled  if  we instead consider 
\be
\bar{D}_a \varphi := D_a \varphi + k D_a \psi \, \varphi, \label{Dbarphi}
\ee
which, thanks to, (\ref{delpsi}) satisfies 
\be
\delta_V (\bar{D}_a \varphi) = (\L_V + k \alpha) \bar{D}_a \varphi .
\ee
Expression (\ref{Dbarphi}) is the desired  definition of  $\diff(S^2)$-covariant derivative for scalars. For arbitrary tensors we can generalize the above reasoning by taking into account the variation of the Christoffel symbols of $D_a$. For example, the $\diff(S^2)$-covariant derivative of a covector $\w_a$ is found to be  given by  
\be
\bar{D}_a \w_b =D_a \w_b  - \Gpsi_{a b}^{c} \, \w_c + k D_a \psi \, \w_b \label{Dawb}
\ee
where
\be
\Gpsi_{a b}^{c}:= -2 D_{(a}\psi \delta_{b)}^c+q_{ab} D^c \psi. \label{Gpsi}
\ee
For general tensors, expression (\ref{Dawb}) generalizes with the appropriate inclusion of $\Gpsi_{a b}^{c}$ symbols  for each tensor index, see Eq. (\ref{Dbq}) for another example.

To summarize, given a general tensor that is covariant with respect to superrotations as in (\ref{delVt}), its $\diff(S^2)$-covariant derivative also transforms covariantly:
\be
\delta_V  \Db_a t_{a_1 \ldots}^{b_1 \ldots} = (\L_V+ k \alpha)t_{a_1 \ldots}^{b_1 \ldots} 
\ee

With the above definitions one can verify  $\bar{D}_a$ satisfies Leibiniz rule with the `weight' $k$ of the product of tensors given by the sum of the weights of each tensor.  

The $\bar{D}_a$ derivative has a number of useful properties we now describe. 

\begin{enumerate}

\item  Its action on $q_{ab}$ is zero:
\be
\Db_{c} q_{ab} =   - \Gpsi_{c a}^{d} q_{d b} - \Gpsi_{c b}^{d} q_{a d} -2 D_c \psi q_{ab} =0. \label{Dbq}
\ee

\item The commutator of $\bar{D}_a$ derivatives satisfies the same formulas as for ordinary covariant derivatives but with a `covariantized' curvature tensor. For instance: 
\be
[\Db_a, \Db_b] \w_c = \Rb_{abc}^{\phantom{abc}d} \w_d \label{commDb}
\ee
where\footnote{In establishing (\ref{defRb})  one needs to use algebraic identities of 2d tensors that may not be manifest in an abstract index notation. These identities are  easily seen in holomorphic coordinates $(z,\zb)$ such that $q_{zz}=q_{\zb \zb}=0$. \label{zzbfnote}}
\be
\Rb_{abcd} = \Rb q_{a[c} q_{d] b}, \quad \text{with} \quad  \Rb= R + 2 D^2 \psi,   \label{defRb}
\ee
$R$ being the scalar curvature of $q_{ab}$. Notice that $ \Rb_{abc}^{\phantom{abc}d}$  is independent of the `weight' $k$ of $\w_c$.

\item Finally, one can show the remarkable property  (see appendix \ref{gerochapp}):
\be
\Db_a \Rb=0 . \label{DbRb} 
\ee
Eq. (\ref{DbRb}) can be thought of as a `covariantized' version of the constant 2d curvature  in Bondi frame. It can also be  understood as a rewriting of the Geroch identity  \cite{geroch} $D_{[a}\rho_{b]c}=0$ where $\rho_{ab} =   \frac{R}{2} q_{ab} - T_{ab}$. See appendix \ref{gerochapp} for further details.
\end{enumerate}

\subsection{Supermomentum revisited}
Let us briefly revisit the expression for supermomentum  in light of the previous discussion. 

We start by noting  that a supertranslation function $f(x)$ should be treated  as a covariant scalar with $k=-1$, since 
$V(f) = \L_V f - \alpha f$.   We can then compute its $\diff(S^2)$-covariant derivative  according to the rules of the previous section. Doing so one finds
\be
-2 \bar{D}_a \bar{D}_b f^\tf = -2 D_a D_b f^\tf  + f T_{ab}. \label{Dbar2f}
\ee
The right hand side of (\ref{Dbar2f})  matches the inhomogeneous term of a  supertranslation (\ref{delf}), and so this identity allows us to reinterpret the expression of  supermomentum  (\ref{PfAS}) as a `$\diff(S^2)$-covariantization' of the  Bondi-frame expression (\ref{delfo}), (\ref{Pbms}).   

In the next sections we use this covariantization idea to extend to non-Bondi frames (i) the zero magnetic condition (\ref{magzero}) and (ii)  the super angular momentum of section \ref{Jbondi}.

\subsection{Asymptotic magnetic condition} \label{magzerosec}

 Give the superrotation transformation rule of $C_{ab}(u,x)$ (\ref{delV}) and the $u \pm \infty$ fall-offs (\ref{fallu}), it is easy to see that   $C_{ab}^\pm(x) = \lim_{u \to \pm \infty} C_{ab}(u,x)$ is a $k=-1$ covariant tensor. Applying the rules of $\Db_a$ differentiation one then finds\footnote{The same comments as those in footnote \ref{zzbfnote} apply here. In the present case, the identity  $D_{[a}\psi D^c C_{b]c} -D^c\psi D_{[a}C_{b]c}=0$ was used to obtain  (\ref{covmag}) .}
\be
 \Db_{[a} \Db^c C^\pm_{b]c}= D_{[a} D^c C^\pm_{b]c} -\frac{1}{2}T_{[a}^c C^\pm_{b]c}. \label{covmag}
\ee
The vanishing of (\ref{covmag}),
\be
 \Db_{[a} \Db^c C^\pm_{b]c}= 0 \label{covmagzero}
\ee
 is the non-Bondi frame generalization of condition (\ref{magzero}) and  imposes that  $C_{ab}^\pm$ is a `pure supertranslation':
 \be 
 C_{ab}^\pm = -2 (\Db_a \Db_b C^{\pm})^\tf  \quad \text{for some function} \quad C^{\pm}. \label{goldstoneC}
 \ee
 where $C^{\pm}$ is a $k=-1$ covariant scalar.\\
 
 \noindent \emph{Comment:} \\
 Equation (\ref{covmag})  illustrates a kind of  complementary for writing expressions, either in terms of $\Db_a$  or in terms of $D_a$  and  $T_{ab}$.  Some properties are more transparent in the first version but obscure in the second version and vice versa.   For instance, the fact that  $C_{ab}^\pm$  in (\ref{goldstoneC}) satisfies (\ref{covmagzero}) is easily seen in the first version, the proof being identical to the one for the round sphere case.  On the other hand,  to see   that  condition  (\ref{covmagzero})   is independent of the $\psi$-ambiguity, we use the second version.

\section{Super angular momentum in a general frame} \label{JVsec}
We now construct the general-frame  candidate for super angular momentum $J_V$ by `covariantizing' the Bondi-frame expression of  section \ref{Jbondi}.   The charge is  a sum of three terms,
\be
J_V = J_V^\hard + J_V^\soft+ J_V^{\partial \I} \label{JV},
\ee
\ba
J_V^\hard & = &  \int_{\I} du d^2 x \sqrt{q} \partial_u C^{ab} \delta_V C_{ab}  \label{Jhard} \\
J_V^\soft & = &    \int_{S^2} d^2 x \sqrt{q} \None^{ab} \sone^V_{ab}  \label{Jsoft} \\
J_V^{\partial \I} & = &   \int_{\partial \I} d^2 x \sqrt{q} (V^a C^{bc} \Db_c C_{ab} + \frac{3}{2}\alphab C^{ab}C_{ab}) \label{Jextra}
\ea
where
\be
\sone^V_{ab} =  \big[ -4 \Db_a \Db_b \alphab + \Db_{(a} \Db^c \delta_V q_{b)c}- \frac{\Rb}{2} \delta_V q_{ab} \big]^\tf , \label{defs}
\ee
 with $\alphab = \Db_a V^a/2 $ ($V^a$ is treated as a $k=0$ vector) and $\Rb$  the `covariantized' scalar curvature defined in (\ref{defRb}).\footnote{The scalar  curvature in the last term term of (\ref{defs}) also appears in  \cite{bms2} and is needed in order for this term to have the same weight as the first two ($k=0$), thus ensuring the correct superrotation transformation properties of $J_V^\soft$,  see  appendix \ref{gbmsaction} .} Note that, as written, it is not obvious that $J_V$ is free from the ambiguity in $\psi$ described in section \ref{psipotential}. We will later give  alternative expressions for $J_V^\soft$ and $J_V^{\partial \I}$ in which this property is manifest.   For now,  let us  focus  on establishing the identity 
\be
\delta_f J_V - P_{V(f)}= \mg(f,V),  \label{delfJPcov}
\ee
where  $\mg(f,V)$ is the covariant version of the Bondi-frame magnetic term (\ref{mag}). This will again vanish due to the zero magnetic condition $\Db_{[a} \Db^c C^\pm_{b]c}=0$.

We start by evaluating the action of a supertranslation on each term of (\ref{JV}). The calculation is essentially the same as that of section \ref{Jbondi} and gives:
\ba
\delta_f J^\hard_V & =  &  P^\hard_{V(f)}  + \int_{S^2} d^2 x \sqrt{q} \Nzero^{ab} (\L_V - \alpha) \szero^f_{ab} \\ 
\delta_f J_V^\soft & = &   - \int_{S^2} d^2 x \sqrt{q} f \Nzero^{ab} \sone^V_{ab} \\
\delta_f J_V^{\partial \I} & = &  \int_{S^2} d^2 x \sqrt{q}  \Nzero^{ab} (-\Db^c(V_a \szero^f_{bc}) +V^c \Db_a  \szero^f_{bc} + 3 \alphab \szero^f_{ab}) , \label{delfJpartial}
\ea
where
\be
\szero^f_{ab} \equiv -2  \Db_a \Db_b f^\tf.
\ee
Let us now look at the contribution in (\ref{delfJPcov}) coming from $\delta_f J^\hard_V - P_{V(f)}$.  The  $P^\hard_{V(f)}$ terms cancel while  the soft terms can be  combined as
\be
(\L_V - \alpha) \szero^f_{ab} - \szero^{V(f)}_{ab}  = \delta_V \szero^f_{ab}. \label{delS0id}
\ee
We thus obtain
\be
\delta_f J^\hard_V - P_{V(f)} =  \int_{S^2} d^2 x \sqrt{q} \Nzero^{ab}  \delta_V \szero^f_{ab}. \label{delfJhard}
\ee
Since the remaining terms  in Eq. (\ref{delfJPcov}) are also proportional to $\Nzero^{ab}$,  it follows that  Eq. (\ref{delfJPcov}) will be satisfied if and only if 
\be
\big[ \delta_V \szero^f_{ab} -f \sone^V_{ab}  -\Db^c(V_{(a} \szero^f_{b)c}) +V^c \Db_{(a}  \szero^f_{b)c} + 3 \alphab \szero^f_{ab} \big]^\tf=
-4 \big[ \Db_{(a} \Db^c(\Db_{[b)}f V_{c]}- \frac{1}{2} f \Db_{[b)}V_{c]})  \big]^\tf. \label{mainid}
\ee
This  identity is equivalent to (the covariant version of)  identity (\ref{Kidentity}) and can be proven  by direct evaluation of both sides, see appendix \ref{Kidapp}.

\subsection{Independence of the ambiguity in $\psi$} \label{psired}
We here verify that $J_V$ depends on $\psi$ only   through $T_{ab}$. 

 The term $J^\hard_V$ (\ref{Jhard}) is independent of $\psi$. 

For $J^\soft_V$ (\ref{Jsoft}) we use the identity:
\be
 \sone^V_{ab}  = \big[  2 \delta_V T_{ab} +  D_{(a} D^c \delta_V q_{b)c}- \frac{R}{2}\delta_V q_{ab}  \big]^\tf \label{S1id}
\ee
which can be established by direct evaluation on both sides, noting that $\bar{\alpha}=- \delta_V \psi$  and taking into account  algebraic 2d identities as in earlier calculations. Eq. (\ref{S1id}) shows that all the $\psi$ dependence of  $(\sone^V_{ab})^\tf$,   and hence of $J^\soft_V$,  is in the term  $\delta_V T_{ab}$ in (\ref{S1id}).

For   $J_V^{\partial \I}$  we use the covariant version of Eq. (\ref{Jextra2}) to rewrite it as
\be
J_V^{\partial \I} =  - \frac{1}{2} \int_{\partial \I} d^2 x \sqrt{q}  C^{ab} (\delta_V \szero^{C}_{ab} -\szero^{\delta_V C}_{ab}-   C \sone^V_{ab}), \label{JVpartial2}
\ee
where  $C|_{\partial \I_{\pm}} \equiv C^{\pm}$ as defined in (\ref{goldstoneC}).   Since the $\psi$ dependence of both $\szero_{ab}$ and $\sone_{ab}$ is through $T_{ab}$ (Eqs. (\ref{Dbar2f}) and (\ref{S1id})), this form makes it manifest that $J_V^{\partial \I}$ is independent  on the $\psi$-ambiguity.

Expression (\ref{JVpartial2}) can also be used to show that   $J_V^{\partial \I}$ vanishes  for global CKVs, since all terms depend  either on $\delta_V q_{ab}$ or on $\delta_V T_{ab}$.  We can see this explicitly by expanding the first two terms in (\ref{JVpartial2}):
\be
 C^{ab}(\delta_V \szero^{C}_{ab} -\szero^{\delta_V C}_{ab}) = C^{ab}(2D^c C D_a \delta_V q_{bc} - D^c C D_c \delta_V q_{ab} +D^2 C \delta_V q_{ab} +C \delta_V T_{ab}), \label{delSzeroC}
\ee
whereas  the last term in (\ref{JVpartial2}) depends on $\delta_V q_{ab}$ and $\delta_V T_{ab}$ according to (\ref{S1id}).\\

\noindent \emph{Comment}: \\
Identities  (\ref{S1id}), (\ref{JVpartial2}) and (\ref{delSzeroC}) show that $J^\soft_V$ and $J_V^{\partial \I}$ depend on $V$ through $\delta_V q_{ab}$ and  $\delta_V T_{ab}$. This property will be crucially used in the next section.

\section{Extension of the gravitational phase space at null infinity} \label{symplectic}
In the previous section we constructed a super angular momentum $J_V$ that is compatible with supertranslations in the sense that 
\be
\delta_f J_V  =  -\delta_V P_f = P_{V(f)} .\label{JVPf}
\ee
On the other hand,  the compatibility of $J_V$ with superrotations
\be
\delta_V J_{V'}  = -J_{[V,V']}, \label{delV0}
\ee
can be  established from the  `superrotation covariance' of the expressions defining $J_V$, see appendix  \ref{gbmsaction}. Finally, compatibility of supermomenta with supertranslations, 
\be
\delta_f P_{f'}=0 \label{delfP0} 
\ee 
is a well  known result that can be easily checked from the expressions of   supertranslations and supermomenta. 

As discussed in the introduction, we can think of  properties (\ref{JVPf}), (\ref{delV0}) and (\ref{delfP0}) as reflecting  an underlying phase space.   In this section we show  these properties can be used to determine an  extension of the  Ashtekar-Streubel  phase space on which   GBMS acts canonically. Let 
\be
\Gamma_{q_{ab}} := \{ C_{ab}(u,x) : q^{ab}C_{ab}=0,  \quad \partial_u C_{ab} \stackrel{u \to \pm \infty}{=} O(1/|u|^{2+\epsilon}), \quad   \Db_{[a} \Db^c C^\pm_{b]c}=0\} ,
\ee
be the space of allowed $C_{ab}$'s for a given  $q_{ab}$. Each $\Gamma_{q_{ab}}$  provides a  realization of the Ashtekar-Streubel  phase space,  with symplectic structure given by 
\be
\Omega_{q_{ab}}= \int_{\I} \sqrt{q}(\delta \partial_u C^{ab} \wedge \delta C_{ab} ), \quad \delta \in \Gamma_{q_{ab}}.  \label{OmegaAS2}
\ee

In the traditional interpretation, different choices of $q_{ab}$ (or frames) are akin to gauge choices. However, to implement superrotations we need to consider the larger  space  \cite{bms2}
\be
\Gamma:= \bigcup_{q_{ab} : \sqrt{q}= \sqrt{\mathring{q}}} \Gamma_{q_{ab}} , \label{defGamma}
\ee 
where $\sqrt{\mathring{q}}$ is the area element of the unit round sphere. 
Our aim is to find  a symplectic structure $\Omega$ on $\Gamma$ such that:
\begin{enumerate}
\item[(i)] $P_f$ and $J_V$ are the canonical charges of  supertranslations and superrotations,
\ba
 \Omega(\delta,\delta_f)  &=& \delta P_f   \quad \forall \delta \in \Gamma \label{delPO}  ,\\
 \Omega(\delta,\delta_V)  & = & \delta J_V    \quad \forall \delta \in \Gamma \label{delJO}, 
\ea
and
\item[(ii)]  $\Omega$ reduces to $\Omega_{q_{ab}}$ when restricted to $\Gamma_{q_{ab}}$,
\be
\Omega|_{\Gamma_{q_{ab}}}=\Omega_{q_{ab}}. \label{omegares}
\ee
\end{enumerate}

Our starting point is to   write  $\Omega$ as
\be
\Omega = \Omega^\I + \Omega^{S^2},
\ee
with  $\Omega^\I $ as in the AS expression (\ref{OmegaAS2})  but allowing for  arbitrary variations in $\Gamma$, 
\be
\Omega^\I  = \int_{\I} \sqrt{q}(\delta \partial_u C^{ab} \wedge \delta C_{ab} ),  \quad \delta \in \Gamma, \label{OI}
\ee
and  $\Omega^{S^2}$ a reminder to be determined. 

Next, we evaluate $\Omega^\I$ on supertranslations and superrotations. A  straightforward calculation gives
\ba
\Omega^\I(\delta,\delta_f) & = & \delta P^\hard_f + \int_{S^2}\sqrt{q} \delta \Nzero^{ab} \szero^f_{ab} \label{OIdelf} \\
\Omega^\I(\delta,\delta_V) & = & \delta J^\hard_V. \label{OIdelV}
\ea
Using  these expressions, conditions   (\ref{delPO}) and (\ref{delJO})  translate into the following conditions on $\Omega^{S^2}$:
\ba
 \Omega^{S^2}(\delta,\delta_f)  &=& \int_{S^2}\sqrt{q}  \Nzero^{ab} \delta \szero^f_{ab}  \label{delPOs} \\
 \Omega^{S^2}(\delta,\delta_V)  & = & \delta J^\soft_V + \delta J^{\partial \I}_V  \label{delJOs}.
\ea

The strategy now will be to  use Eq.  (\ref{delJOs}) to determine $\Omega^{S^2}$ and then verify Eq.  (\ref{delPOs}).   

We  assume   $\Omega^{S^2}$ is of the form
\be
\Omega^{S^2}  =  \O^\soft+\O^{\partial \I}, \label{Osum} 
\ee
with
\be
\O^{\cdots}(\delta,\delta_V)  =  \delta J^{\cdots}_V , \label{OdotsJ}
\ee
where ``$\cdots$'' stands for either ``soft'' or ``$\partial \I$''. We think of each $\O^{\cdots}$ as defining a  symplectic structure on its own such that  $\delta_V$ acts canonically  with charge $J^{\cdots}_V$. 

It will  be convenient to express  $\O^{\cdots}$ in terms of a symplectic potential $\theta^{\cdots}$,
\be
\O^{\cdots}(\delta,\delta') = \delta \theta^{\cdots} (\delta') - \delta' \theta^{\cdots}(\delta) - \theta^{\cdots}([\delta,\delta']),
\ee
and  consider a $\theta^{\cdots}$   compatible with $\delta_V$  (see e.g.  \cite{abr})
\be
  \theta^{\cdots}(\delta_V) =  J^{\cdots}_V ,\label{thetaJ}
\ee
so that  Eq.  (\ref{OdotsJ}) becomes\footnote{Eq. (\ref{lietheta})  can  be thought of as a $\delta_V$-invariance  condition on $\theta^{\cdots}$ \cite{abr}.}
\be
\delta_V \theta^{\cdots}(\delta) + \theta^{\cdots}([\delta,\delta_V]) = 0. \label{lietheta}
\ee

If we now look at the expressions for $J^\soft_V$ and $J^{\partial \I}_V$  as given by Eqs. (\ref{Jsoft}),  (\ref{S1id}) and (\ref{JVpartial2}),   we can easily find candidates for  $\theta^{\cdots}(\delta)$  satisfying (\ref{thetaJ}) by doing the replacement $\delta_V \to \delta$ in such expressions. Defining
\be
\sone_{ab}(\delta) :=  \big[ 2 \delta T_{ab} +  D_{(a} D^c \delta q_{b)c}- \frac{R}{2}\delta q_{ab}  \big]^\tf
\ee
so that $\sone_{ab}(\delta_V) = \sone_{ab}^V$, we find  the following candidates for  symplectic potentials: 
  \ba
\theta^\soft(\delta) &=& \int_{S^2} \sqrt{q} \None^{ab} \sone_{ab}(\delta), \label{thetasoft}\\
\theta^{\partial \I}(\delta) &=&  - \frac{1}{2} \int_{\partial \I}  \sqrt{q}  C^{ab} (\delta \szero^{C}_{ab} -\szero^{\delta C}_{ab}-   C \sone_{ab}(\delta)). \label{thetaI}
\ea
By construction, (\ref{thetasoft}) and (\ref{thetaI}) satisfy Eq. (\ref{thetaJ}). Condition (\ref{lietheta}) can then be shown to be a  a consequence of (i) the fact that $\theta^{\cdots}(\delta)$ is only sensitive to variations  of $q_{ab}$, (ii) the fact that any variation $\delta q_{ab}$ can be written as $\delta q_{ab}=\delta_W q_{ab}$ for some vector field $W^a$ and (iii) the `superrotation covariance' of $J^{\cdots}_V$. Indeed,  recall that $T_{ab}$ is fully determined by $q_{ab}$ and  note that the term $\delta \szero^{C}_{ab} -\szero^{\delta C}_{ab}$ is independent of variations of $C$. Writing  $\delta q_{ab}= \delta_{W}q_{ab}$  condition  (\ref{lietheta}) becomes 
\be
\delta_V \theta^{\cdots}(\delta_{W}) + \theta^{\cdots}([\delta_{W},\delta_V]) = 0  \label{lietheta2}.
\ee
 Using $[\delta_{W},\delta_V]= - \delta_{[W,V]}$ \cite{bms2} and (\ref{thetaJ}), Eq. (\ref{lietheta2}) can be seen to be a direct consequence of the superrotation covariance of $J^{\cdots}_{W}$ (see appendix \ref{gbmsaction}),
 \be
 \delta_V J^{\cdots}_{W} = -J^{\cdots}_{[V,W]}.
 \ee 

Summing the resulting  $\Omega^\soft$ and  $\Omega^{\partial \I}$, we obtain    $\Omega^{S^2}$ satisfying  (\ref{delJOs}).  It now remains to verify that such $\Omega^{S^2}$ satisfies  (\ref{delPOs}). This can be shown to be a consequence of the supertranslation transformation properties  of $J^\hard_{V}$ and $J^{\cdots}_V$ as follows. 

Written in terms of the symplectic potential  $\theta^{S^2}= \theta^\soft+\theta^{\partial \I}$,  condition  (\ref{delPOs})  takes the form,
\be
\delta \theta^{S^2} (\delta_f) - \delta_f \theta^{S^2}(\delta) - \theta^{S^2}([\delta,\delta_f]) = \int_{S^2}\sqrt{q}  \Nzero^{ab} \delta \szero^f_{ab} . \label{delPOs2}
\ee

As before, we note that Eq. (\ref{delPOs2}) is only sensitive to variations  of $q_{ab}$. Writing $\delta q_{ab}= \delta_W q_{ab}$ for some $W^a$ and using the fact that $\theta^{S^2}$ vanishes if evaluated on  variations with $\delta q_{ab}=0$ (first and third term in (\ref{delPOs2})), the condition reduces to
\be
-\delta_f \theta^{S^2}(\delta_W)  = \int_{S^2}\sqrt{q}  \Nzero^{ab} \delta_W \szero^f_{ab} ,
\ee 
or, using (\ref{thetaJ}), to
\be
- \delta_f ( J^\soft_W +  J^{\partial \I}_W ) =  \int_{S^2}\sqrt{q}  \Nzero^{ab} \delta_W \szero^f_{ab} .\label{delPOs4}
\ee 
By writing  $-(J^\soft_W +  J^{\partial \I}_W)= (J_W^\hard-J_W)$, Eq. (\ref{delPOs4})  can be seen to be a consequence of the supertranslation transformation formulas (\ref{delfJhard}) and (\ref{JVPf})   for  $J^\hard_W$ and $J_W$ respectively.

\subsection{Summary}
We  found a symplectic structure $\Omega$ on the space  $\Gamma$  (\ref{defGamma}) satisfying  Eqs. (\ref{delPO}) and (\ref{delJO}). $\Omega$ is written as a sum of  `bulk'  and  `boundary' pieces:
\be
\Omega = \Omega^\I + \Omega^{S^2},
\ee
where $\Omega^\I$ (\ref{OI}) is the extension  to $\Gamma$ of the AS symplectic structure  and $\Omega^{S^2}$ is given in terms of a symplectic potential 
\be
\theta^{S^2} = \theta^\soft+ \theta^{\partial \I},
\ee
with $\theta^\soft$ and $\theta^{\partial \I}$ defined in Eqs. (\ref{thetasoft}) and (\ref{thetaI}) respectively.  

By integration by parts on the sphere, one can bring $\theta^{S^2}$  and $\Omega^{S^2}$ into a form
\ba
\theta^{S^2} & =& \int_{S^2}\sqrt{q}( p^{ab} \delta q_{ab} + \Pi^{ab} \delta T_{ab} ) \\
 \Omega^{S^2}  &= &\int_{S^2}\sqrt{q}( \delta p^{ab} \wedge \delta q_{ab} + \delta \Pi^{ab} \wedge \delta T_{ab} )
\ea
where
\ba
p^{ab} &= & D^{(a} D_c \None^{b) c} - \frac{R}{2} \None^{ab}  + (\text{quadratic in $C$})^{ab}|_{\partial \I} \label{pab} \\
\Pi^{ab} &=& 2 \None^{ab}+ \frac{1}{2} C C^{ab}|_{\partial \I}. \label{Piab}
\ea
The terms quadratic in $C^{\pm}$ in (\ref{pab}) can be obtained with the help of  Eq. (\ref{delSzeroC}).  Written in this form, it is clear that $ \Omega^{S^2}|_{\Gamma_{q_{ab}}}=0 $ and so condition  (\ref{omegares}) is also satisfied.

The terms linear in $\None^{ab}$ in (\ref{pab})  correspond to those found in \cite{bms2,compere} by covariant phase space methods.  An important question we leave open is whether the  full $\Omega$ can be understood from a covariant phase space perspective.   In this respect, we note the recent work  \cite{flanagan}  which shows there cannot be a symplectic structure --constructed from a local and covariant symplectic current-- at null infinity  that supports the action of GBMS.  There is in principle no contradiction with our results, since we do not assume a symplectic current and furthermore  our symplectic structure  contains non-local terms due to appearance of  $T_{ab}$ (which depends non-locally on $q_{ab}$) and of $C^{\pm}$ (which depends non-locally on $C^{\pm}_{ab}$).  It would be interesting to see if the analysis of \cite{flanagan} can be extended  to include such  non-local terms. \\

We conclude  by noting that conditions (\ref{delPO}), (\ref{delJO}) together with (\ref{JVPf}), (\ref{delV0}) and (\ref{delfP0}) imply the PBs
\be
\{P_{f}, P_{f'}\}=0, \quad \{J_V,P_f\} = P_{V(f)}, \quad   \{ J_V,J_{V'} \} = J_{[V, V']}  .
\ee

\section{Discussion} \label{finalsec} 
If the symmetries of gravity at null infinity are to include all  diffeomorphisms on the celestial sphere \cite{bms1}, one should be able to associate canonical charges to this $\diff(S^2)$ group.   A prerequisite for this is to have a phase space on which these symmetries act. It is clear that such phase space should include, in addition to gravitational radiation,  the `kinematical' 2-metric  at null infinity \cite{bms2}. Doing so, however,   introduces several divergences in the symplectic structure at null infinity that are notoriously difficult to control  \cite{bms2,compere,flanagan}.  

Here we have taken an alternative route.  Rather than trying to obtain a finite symplectic structure from the  beginning, we started by noting certain conditions the GBMS charges should satisfy if such finite symplectic structure exists. In particular, we noted that the  $\diff(S^2)$ transformation properties of supermomenta imply there should \emph{not} be an extension term in the Poisson bracket between supermomenta and super angular momenta.\footnote{We remind the reader that our statements refer to  generators on the phase space associated to the entire  null infinity. We are not making any claim about  surface charges --as  those studied in \cite{BTcharge,compere}--   associated to finite cuts of null infinity.} 
This led us  to consider a correction term in the expression of super angular momentum that  cancels (a non-CKV analogue of) the BT extension term \cite{BTcharge}.  The corrected super angular momentum may offer a better understanding of the charge algebra at the level of the gravitational $S$ matrix \cite{extradouble}: The BT extension  does not commute with the $S$ matrix \cite{dfh} and hence  appears to contradict the idea of (extended/generalized) BMS as a symmetry of gravitational scattering \cite{strom0,stromST,stromvirasoro}.

We finally showed there exists a natural symplectic structure  at null infinity that is compatible with the expressions of GBMS charges described above. A by product of our analysis was the introduction of a $\diff(S^2)$-covariant derivative at null infinity that  may be of interest beyond the scope of this paper. 

There are many directions this work should be improved  upon. Whereas the finding of a symplectic structure at null infinity supporting the $\diff(S^2)$ action is a non-trivial fact  --there was no guarantee of its existence-- the question remains open as to  whether this structure can be obtained from covariant phase space methods.   A related question is whether there can be a 3+1 realization of GBMS as was recently shown to exist for BMS \cite{henneaux}. 

We have worked under the assumption of `tree-level' $u$ fall-offs in which the news tensor decays faster than $1/u^2$. However, in generic gravitational scattering the news tensor has a leading component falling exactly as $1/u^2$ \cite{laddhasen,sahoosen} that should be incorporated in the analysis. 
 
 The GBMS group has a direct analogue in higher dimensions and one may  ask if the results presented here can be generalized to higher dimensions as in the BMS case  \cite{hd1,hd2,hd3}.  Finally, one may wonder if there could exist an additional  extension of the gravitational phase space that supports  large diffeomorphisms \cite{subsub1,subsub2} associated to the sub-subleading  soft graviton theorem \cite{stromcach,subsubls}.

\begin{acknowledgments}
This project grew out from discussions with Alok Laddha. We would like to thank him  for all his feedback,  guidance, and encouragement.  We are also grateful to Rodrigo Eyheralde, Rodolfo Gambini, Guzman Hernandez-Chifflet,  Michael Reisenberger and Aureliano Skirzewski  for helpful discussions. MC would like to thank Alok Laddha and Raphael Flauger for illuminating  discussions on superrotation charges during the MIAPP Program ``Precision Gravity: From the LHC to LISA''.  We are grateful to Biswajit Sahoo for corrections made on the first version of the manuscript. This research was supported in part by the Munich Institute for Astro- and Particle Physics (MIAPP) which is funded by the Deutsche Forschungsgemeinschaft (DFG, German Research Foundation) under Germany's Excellence Strategy – EXC-2094 – 390783311.
\end{acknowledgments}

\appendix
\section{Finite $\diff(S^2) \subset \gbms$ transformations} \label{diffeos}
In this appendix, following \cite{compere,comperelong}, we calculate the action of finite superrotations $\phi \in \diff(S^2)$ on the asymptotic spacetime metric. Restricting to the  case where the `initial' spacetime metric is in a  Bondi-frame, we will obtain a parametrization of  various non-Bondi frame quantities in terms of  $\phi$. 

The idea is to proceed in the same way as for infinitesimal superrotations \cite{BT,bms2} but for finite diffeos. Namely, we look for  spacetime diffeomorphisms that preserve the space-time metric in Bondi gauge,
\be
g_{rr}=0=g_{ra}, \quad \sqrt{\det g_{ab}} = r^2 \sqrt{q}, \label{bondigge}
\ee
with the   standard $1/r$ fall-offs \cite{BT} . We express the  diffeo as\footnote{In this appendix  $R$ denotes a radial coordinate and $\R$ the scalar curvature of $q_{ab}$. In the rest of paper $R$ is used for the scalar curvature of $q_{ab}$.}
\be
(r,u,x^a) \to (R,U,X^A)
\ee 
and assume an $1/r$ expansion  compatible with that of the spacetime metric:
\ba
R(r,u,x)  &=& \Rzero(x) r + \Rone(u,x)  +O(1/r) \label{R} \\
U(r,u,x)  &=& \Rzero^{-1}(x) u  + O(1/r) \label{U} \\
X^A(r,u,x) & =& \phi^A(x) + \frac{1}{r}\Xone^A(u,x) +O(r^{-2}). \label{XA}
\ea
In the above expressions we have already fixed some of the $u$-dependence that is required for compatibility with the Bondi metric \cite{BT,bms2}. We have also  excluded supertranslations, which correspond to  a $u$-independent term in the $O(r^0)$ part of $U$ (see \cite{comperelong} for expressions of finite supertranslations). 

We proceed as follows. The `initial' spacetime metric in $(R,U,X^A)$ coordinates is  taken to be in Bondi frame so that its angular components take the form
\be
g_{AB}(R,U,X)= R^2 q_{AB}+ R C_{AB}(U,X) + \cdots 
\ee 
with $q_{AB}$ the unit round sphere metric. Next, we compute the various components of the pullback metric in the $(r,u,x^a)$ coordinates under the spacetime diffeo  (\ref{R}, \ref{U}, \ref{XA}). By imposing the  Bondi gauge conditions on the pullback  metric we then  determine the spacetime diffeo coefficients in terms of $\phi \in \diff(S^2)$.  

The angular part of the pullback metric is found to be given by
\be
g_{ab}(r,u,x) = r^2 q^\phi_{ab}(x) + r C^\phi_{ab}(u,x) + \cdots
\ee
where
\be
 q^\phi_{ab}(x) = \Rzero^2 \partial_a \phi^A \partial_b \phi^B q_{AB}(\phi), \label{qphi}
 \ee
 and
 \begin{multline}
 C^\phi_{ab}(u,x)  = \Rzero \partial_a \phi^A \partial_b \phi^B\big(C_{AB}(\Rzero^{-1} u, \phi)+2 \Rone  q_{AB}(\phi) + \Rzero \Xone^C \partial_C q_{AB}(\phi) \big)  \\ + 2 \Rzero^2 \partial_a \phi^A \partial_b \Xone^B q_{AB}(\phi) + 2 \Rzero^{-2}\partial_a \Rzero \partial_b \Rzero. \label{Cphi}
\end{multline}
In the RHS of the above equations it is  understood that $\phi, \Rzero, \Rone$, etc. are evaluated at $(u,x)$ as in 
Eqs. (\ref{R}, \ref{U}, \ref{XA}). 

To leading order, the determinant condition  (\ref{bondigge}) implies $\det q^\phi(x)=\det q(x)$, which fixes   $\Rzero(x)$ to be
\be
\Rzero(x) = \frac{\det^{1/4}q(x)}{\det^{1/4}q(\phi(x))}\frac{1}{ \det^{1/2} \partial \phi(x)}. \label{Rzero}
\ee
To subleading order, the determinant condition implies
\be
q_\phi^{ab} C^\phi_{ab}=0,
\ee
which fixes $\Rone$ in terms of the other quantities. The remaining diffeo component to be determined  is $\Xone^A$, which can be obtained from condition $g_{ra}=0$. The pullback  for such metric components is found to be
\be
g_{ra}(r,u,x) = u \partial_a \ln \Rzero - \Xone^A \partial_a \phi^B \Rzero^2 q_{AB}(\phi) + O(r^{-1}). \label{gra}
\ee
To solve $g_{ra}=0$ it is convenient to write  $\Xone^A$ as a pushforward of a vector field $Y^a$,
\be
\Xone^A = \partial_a \phi^A Y^a.
\ee
Eq. (\ref{gra}) can then be written as
\be
g_{ra} = u \partial_a \ln \Rzero - q^\phi_{ab}Y^b  + O(r^{-1}),
\ee
from which we obtain
\be
Y^a = u q_\phi^{ab} \partial_a \ln \Rzero.
\ee
We now have all elements to express $C^\phi_{ab}$ in (\ref{Cphi}) in terms of $\phi$. The expression simplifies when written in terms of the covariant derivative $D^\phi_a$  compatible with the metric $q^\phi_{ab}$. After some work, it can be expressed as  
\be
C^\phi_{ab}(u,x)  = \Rzero \partial_a \phi^A \partial_b \phi^B C_{AB}(\Rzero^{-1} u, \phi) + u \Nvac_{ab} \label{Cphi2}
\ee
where
\be
\Nvac_{ab}= 2 (D^\phi_a \ln \Rzero D^\phi_b \ln \Rzero + D^\phi_a D^\phi_b \ln \Rzero)^\tf. \label{Nvac}
\ee
This result is essentially that of section 3 of \cite{compere}, except that here our `initial' 2d metric is the round unit sphere, whereas in \cite{compere} it is the Euclidean plane.  
 
The  above expressions can be used to identify the potential $\psi$ of section \ref{psipotential} in terms of $\phi$.  From  (\ref{Cphi2}) and (\ref{Nvac}) we see that
\be
T_{ab}= \Nvac_{ab}, \quad  \psi = \ln \Rzero, \label{psiRzero}
\ee
where  $\Rzero$ in terms of $\phi$ is given in Eq. (\ref{Rzero}). 

After this identification, we can write Eq. (\ref{qphi}) as
\be
 q^\phi_{ab}(x) = e^{2 \psi} \partial_a \phi^A \partial_b \phi^B q_{AB}(\phi). \label{qphi2}
 \ee
From this perspective,  $\psi$ appears as  a conformal rescaling that makes $q_{ab}$   diffeomorphic to the unit sphere metric.   We finally use Eq. (\ref{qphi2}) 
to obtain a formula for the scalar curvature $\R$ of $ q^\phi_{ab}$ in terms of $\psi$:
\be
\R = 2(e^{-2 \psi} - (D^\phi)^2 \psi). \label{Rpsi}
\ee
This is the analogue of Eq. (3.11) in \cite{compere}.

\section{Geroch tensor} \label{gerochapp}
In the Geroch approach \cite{geroch}, $T_{ab}$ is (minus) the trace-free part of a tensor
\be
\rho_{ab} = \frac{R}{2} q_{ab} - T_{ab}, \label{defrho}
\ee
that satisfies 
\be
D_{[a}\rho_{b]c}=0.  \label{gerocheq}
\ee
Geroch shows there is a unique tensor $\rho_{ab}$ satisfying the above conditions. We here verify our  $T_{ab}$ satisfies Eq.  (\ref{gerocheq}). 

Inserting \ref{defrho} in \ref{gerocheq}, the Geroch condition becomes
\be
D_{[a}R q_{b]c}= 2 D_{[a}T_{b]c} \label{geroch2}.
\ee
Using 2d algebraic relations, this can  be shown to be equivalent to\footnote{As for similar identities used in the paper, this equivalence is easily seen  in holomorphic coordinates.}
\be
D_a R = -2 D^b T_{ab} .\label{geroch3}
\ee
Eq. (\ref{geroch3}) corresponds to Eq. (3.12) of \cite{compere} and can be  shown to follow from the expressions of $R$ (\ref{Rpsi}) and $T_{ab}$  (\ref{defT}) in terms of $\psi$:
\ba
R &= &2(e^{-2 \psi} - D^2 \psi) \label{Rpsi2} \\
T_{ab}& =& 2 (D_a \psi D_b \psi + D_a D_b \psi)^\tf.
\ea 

We conclude by discussing the implications of these identities on  the 
`covariantized' scalar curvature $\Rb$. We first note that  $\Rb$ given in Eq. (\ref{defRb})  can be rewritten, using Eq. (\ref{Rpsi2}), as:
\be
\Rb = 2 e^{-2 \psi}.
\ee 
In this form, the superrotation covariance of $\Rb$ appears as a direct consequence of the transformation property of $\psi$ (\ref{delpsi}), from which one obtains
\be
\delta_V \Rb = \L_V \Rb + 2 \alpha \Rb.
\ee
Finally, if we compute $\Db_a \Rb$ with the rules of section \ref{covderivative} one can verify 
\be
\Db_a \Rb = D_a R +2 D^b T_{ab},
\ee
which vanishes due to the Geroch identity (\ref{geroch3}).

\section{Main identity} \label{Kidapp}
Let us first recall that  $\delta_f J_V^{\partial \I}$ and  $\mg(f,V)$ can be written in terms of $ \Nzero^{ab}$ as
\ba
\delta_f J_V^{\partial \I} & =&   \int_{S^2}  \sqrt{q}  \Nzero^{ab} B_{ab}  \\
 \mg(f,V) &= & \int_{S^2}  \sqrt{q}  \Nzero^{ab}  M_{ab}.
\ea
where
\ba
B_{ab} & = &\big[ -\Db^c(V_{a} \szero^f_{b c}) +V^c \Db_{a}  \szero^f_{b c} + 3 \alphab \szero^f_{ab} \big]^\stf , \label{defB}\\
M_{ab} & = &   \big[  \Db_{a} \Db^c \left(-2 \Db_{b}f V_{c}  + 2  \Db_{c}f   V_{b}   +  f \Db_{b}V_{c} -  f \Db_{c}V_{b}\right)  \big]^\stf, \label{defM}
\ea
where STF stands for symmetric, trace-free part in the indices $(a,b)$.

Equation  (\ref{delfJPcov}) 
\be
\delta_f J_V - P_{V(f)}= \mg(f,V)  \label{delfJPcovapp}
\ee
then becomes
\be
\int_{S^2}  \sqrt{q}  \Nzero^{ab} (\delta_V \szero^f_{ab} -f \sone^V_{ab} +B_{ab}) =  \int_{S^2}  \sqrt{q}  \Nzero^{ab} M_{ab}
\ee
which is satisfied if and only if  
\be
\big[ \delta_V \szero^f_{ab}\big]^\tf  -f \sone^V_{ab}  + B_{ab} = M_{ab}. \label{mainidapp}
\ee
This is the identity we wish to prove, which corresponds to Eq. (\ref{mainid}) of section \ref{JVsec}.\footnote{Alternatively, in the version of section \ref{Jbondi}, Eq. (\ref{Kidentity}) corresponds to $\big[ \delta_V \szero^f_{ab}\big]^\tf -f \sone^V_{ab} =-  B_{ab} + M_{ab}$.}

We start by evaluating $B_{ab}$. Expanding the derivatives in (\ref{defB})  and using the properties listed at the end of section \ref{covderivative}  one finds
\be
B_{ab} = \big[ 2 \Db^c V_a \Db_b \Db_c f + \Rb V_a D_b f+2 V_a \Db_b \Db^2 f- \frac{1}{2} \gamma_{ab} \Db^2 f -2 V^c \Db_a \Db_b \Db_c f - 6 \alphab \Db_a \Db_b f \big]^\stf \label{expandedB}
\ee
where we have introduced the notation,
\be
\gamma_{ab}=  2 \big[ \Db_a V_b \big]^\stf = \delta_V q_{ab}. \label{defgamma}
\ee
Notice that the second equality in (\ref{defgamma}) involves a  non-trivial identity  $ \big[ \Db_a V_b \big]^\stf =  \big[ D_a V_b \big]^\stf$. We will use the notation (\ref{defgamma}) in all remaining expressions.

Expanding now (\ref{defM}) and comparing with (\ref{expandedB}) one finds
\begin{multline}
B_{ab} - M_{ab}  = f \big[ -2 \Db_a \Db_b \alphab + \Db_a \Db^c \gamma_{bc}- \frac{\Rb}{2} \gamma_{ab} - \Db_a \Rb V_b \big]^{\stf}   \\ 
+\big[ \Db_a f \Db^c f \gamma_{bc} - \Db^c f \Db_a \gamma_{bc} + \Db_a \Db^c f \gamma_{bc} - \frac{3}{2} \gamma_{ab}\Db^2 f    \big]^{\stf}, \label{BminusM}
\end{multline}
where we note that the last term in the first line is actually zero since   $\Db_a \Rb =0$. 

Finally, we evaluate the first term in (\ref{mainidapp}) 
\be
\big[ \delta_V \szero^f_{ab} \big]^\tf = \big[- 2 f \Db_a \Db_b \alphab +  2 \Db^c f \Db_a \gamma_{bc} - \Db^c f \Db_c \gamma_{ab} +\gamma_{ab} \Db^2 f   \big]^{\stf}, \label{delVSzero}
\ee
and recall the definition of $\sone^V_{ab}$ (\ref{defs})
\be
\sone^V_{ab} = \big[  -4 \Db_a \Db_b \alphab + \Db_{a} \Db^c \gamma_{b c}- \frac{\Rb}{2} \gamma_{ab} \big]^\stf . \label{SoneV}
\ee

Collecting (\ref{BminusM}), (\ref{delVSzero}) and (\ref{SoneV}) we  finally arrive at
\begin{multline}
\big[ \delta_V \szero^f_{ab}\big]^{\stf}  -f \sone^V_{ab}  + B_{ab} - M_{ab} = \\
\big[\Db^c f \Db_a \gamma_{bc} - \Db^c f \Db_c \gamma_{ab} + \Db_a f \Db^c \gamma_{bc} + \Db_a \Db^c f \gamma_{bc} - \frac{1}{2} \gamma_{ab} \Db^2 f \big]^\stf. \label{mainidfinal}
\end{multline}
One can now check that the right hand side of (\ref{mainidfinal}) is  trivially zero. As for similar 2d algebraic identities,  this can be easily seen by writing the expression in holomorphic coordinates.  This concludes the proof of identity (\ref{mainidapp}) and hence of (\ref{delfJPcovapp}).

\section{Superrotation covariance of charges} \label{gbmsaction}
In this appendix  we show the relations
\ba
\delta_V P_f & = & - P_{V(f)} \label{delVP2} \\
\delta_V J_{V'} & =& -J_{[V,V']}, \label{delV2}
\ea
which express  the superrotation covariance of charges.

Let us first verify the `hard' parts of (\ref{delVP2}), (\ref{delV2}). For the supermomentum we have
\ba
\delta_V P^\hard_f & = & \int_{\I}\sqrt{q} f \left(\partial_u \delta_V  C^{ab}  \partial_u C_{ab} + \partial_u C^{ab} \partial_u \delta_V C_{ab} \right) \\
& = & \int_{\I} \sqrt{q} f \left( (\L_V \dot{C}^{ab}+\alpha u \ddot{C}^{ab}+ 4 \alpha \dot{C}^{ab}) \dot{C}_{ab}+  \dot{C}^{ab}(\L_V \dot{C}_{ab}+\alpha u \ddot{C}_{ab}) \right) \\
& = & \int_{\I} \sqrt{q} f ( \L_V \dot{C}^2+\alpha u \partial_u \dot{C}^2 + 4 \alpha \dot{C}^2 ) \\
& = & \int_{\I} \sqrt{q} (-\L_V f \dot{C}^2+f \alpha \dot{C}^2 ) = -P^\hard_{V(f)},
\ea
where in the last step we integrated by parts and used that $\L_V \sqrt{q} = 2 \alpha \sqrt{q}$. To simplify notation we denote $u$-derivatives with dots and $\dot{C}^2 \equiv \dot{C}^{ab}\dot{C}_{ab}$.

Similarly, for the super angular momentum we have
\ba
\delta_V J^\hard_{V'} & = & \int_{\I}\sqrt{q} (\partial_u \delta_V C^{ab} \delta_{V'}C_{ab}+ \dot{C}^{ab} \delta_V \delta_{V'} C_{ab} \\
& = & \int_{\I}\sqrt{q}  \left( (\L_V \dot{C}^{ab}+\alpha u \ddot{C}^{ab}+ 4 \alpha \dot{C}^{ab}) \delta_{V'}C_{ab}+ \dot{C}^{ab}\delta_V \delta_{V'} C_{ab} \right) \\
& = & \int_{\I}\sqrt{q}  \left( \dot{C}^{ab}(-\L_V \delta_{V'}C_{ab} - \alpha u \partial_u  \delta_{V'}C_{ab} + \alpha \delta_{V'}C_{ab}) + \dot{C}^{ab}\delta_V \delta_{V'} C_{ab} \right) \\
& = &  \int_{\I}\sqrt{q}\dot{C}^{ab} ( -\delta_{V'} \delta_V C_{ab} +\delta_V \delta_{V'} C_{ab}) =-J^\hard_{[V,V']},
\ea
where in the third step we integrated by parts and in the fourth we recognized the combination $-\delta_{V'}\delta_V C_{ab}$ in the first term of the third line. Finally we used $[\delta_V,\delta_{V'}] C_{ab}= - \delta_{[V,V']}C_{ab}$ \cite{bms2}. 

We now discuss the remaining, $u$-independent terms of the charges.  We will  use  the notion of superrotation covariance of section \ref{covderivative} to facilitate the calculation. Let us start by noting that  a `covariant' scalar $\rho(x)$ with $k=+2$ has a superrotation invariant integral over the sphere,
\ba
\delta_V \int_{S^2}  \sqrt{q} \rho  &=&  \int_{S^2} d^2 x \sqrt{q} \delta_V \rho \\
 &=&  \int_{S^2}  \sqrt{q} (\L_V \rho + 2 \alpha \rho) =0
\ea
where in the last equality we integrated by parts and  used $\L_V \sqrt{q} = 2 \alpha \sqrt{q}$.  Next, we note that eventhough $f$ and ${V'}^a$ are parameters and hence do not transform under the  action of $\delta_V$, they can be thought of as `covariant'  tensors with $k=-1$ and $k=0$ respectively, due to the GBMS algebra  relations (\ref{bmsalgebra0}).  One can then check that all the integrands of  $P_{f}^\soft$, $J^\soft_{V'}$ and $J^{\partial \I}_{V'}$ have $k=2$. However because $f$ and $V'^a$ are parameters that do not change under $\delta_V$ one gets
\ba
\delta_V P^\soft_f & = & - P^\soft_{V(f)} \label{delVPsoft}  \\
\delta_V J^\soft_{V'} & =& -J^\soft_{[V,V']} \\
\delta_V J^{\partial \I}_{V'} & =& -J^{\partial \I}_{[V,V']} 
\ea
rather than zero. 

Let us do the calculation in detail for (\ref{delVPsoft}), the others following along the same lines:
\ba
\delta_V \int_{S^2} \sqrt{q} \Nzero^{ab}\szero^f_{ab} & = & \int_{S^2} \sqrt{q}(\delta_V \Nzero^{ab}\szero^f_{ab} +\Nzero^{ab}\delta_V \szero^f_{ab} ) \\
& = & \int_{S^2} \sqrt{q}(- \Nzero^{ab}(\L_V -\alpha)\szero^f_{ab} +\Nzero^{ab}\delta_V \szero^f_{ab} ) \\
& = & - \int_{S^2} \sqrt{q} \szero^{V(f)}_{ab} = - P^\soft_{V(f)} ,
\ea
where in the second line we integrated by parts on the sphere and in the last line we used Eq. (\ref{delS0id}).


\end{document}